\documentclass[prd,preprintnumbers,showpacs,onecolumn,showkeys,floatfix,nofootinbib,notitlepage]{revtex4-1}
\usepackage{graphicx}
\usepackage{amssymb}
\usepackage{amsmath}
\usepackage{mathtools}
\usepackage{epsfig}
\usepackage{color}
\usepackage{enumerate}
\usepackage{slashed}
\usepackage{hyperref}
\usepackage{epstopdf}

\def\slashchar#1{\setbox0=\hbox{$#1$}
  \dimen0=\wd0 \setbox1=\hbox{/} \dimen1=\wd1
   \ifdim\dimen0\big>\dimen1 \rlap{\hbox to \dimen0{\hfil/\hfil}} #1
   \else  \rlap{\hbox to \dimen1{\hfil$#1$\hfil}} / \fi}

\newcommand{\ud}{\mathrm{d}}
\newcommand{\be}{\begin{equation}}
\newcommand{\ee}{\end{equation}}
\newcommand{\bea}{\begin{eqnarray}}
\newcommand{\eea}{\end{eqnarray}}

\newcommand{\Appendix}[1]%
   {%
     \section{#1}%
      }

\begin{document}

\title{Cancellation of Infrared Divergence in Inclusive Production of Lepton Pair Near the Threshold of Heavy Quarkonia }

\author{Gao-Liang Zhou
}
\affiliation{Center for High Energy Physics, Peking University,  Beijing 100190, People's Republic of China}




\begin{abstract}
The detailed proof of cancellation of topologically unfactorized infrared divergences in the inclusive production of lepton pair close to the threshold of heavy quarkonia is presented. To make the effects of  transition between states containing heavy quark pairs, which is important in such cancellation, the final detected states are constrained to be lepton pair near the threshold of the heavy quarkonia instead of heavy quarkonia themselves. Such cancellation is crucial for the NRQCD factorization of these processes.
\end{abstract}

\maketitle

\section{Introduction}
\label{introduction}

In spite of the great importance of  Non-relativistic QCD(NRQCD) factorization in study of processes involving heavy quarkonia(see, for example, Refs.
\cite{Brambilla:2010cs,BBEOP2013,Braaten:1994vv,Cho:1995vh,Cho:1995ce,Amundson:1996ik,Beneke:1996yw,Braaten:1996jt,Cacciari:1996dg,
Huang:1996cs,Braaten:1999qk,Braaten:2000cm,Gong:2007db,Ma:2010yw,Butenschoen:2011yh,Gong:2013qka}), such factorization theorem  for inclusive production of heavy quarkonia remains a non-trivial issue(\cite{NQS:2005,Nayak:2006fm,zhou:2015inc}). In the original work on NRQCD(\cite{Caswell:1985ui,Bodwin:1994jh}), the argument about the NRQCD factorization for these process rely on the KLN(\cite{KLN:1962,KLN:1964}) cancellations of infrared divergences. In \cite{Bodwin:1994jh}, infrared divergences caused by gluons exchanged between heavy quarks and other states are believed to be cancel out once the summation over undetected states has been made. Infrared divergences caused by gluons exchanged between the heavy quark pair are absorbed into NRQCD matrix elements. NRQCD factorization theorem can be proved steadily in exclusive  production of two charmonia in $e^{+}e^{-}$ annihilation and production of a charmonium and a light meson in B-meson decays.(\cite{Bodwin:2008nf,Bodwin:2010fi}). For inclusive processes, however, authors in \cite{NQS:2005,Nayak:2006fm} discover that  infrared divergences caused by gluons exchanged between  heavy quarks and other states do not cancel out at next to next leading order(NNLO) in $\alpha_{s}$ once the final detected states are constrained to be color singlet heavy quark pair.  These divergences are not topologically factorized. According to these results, it is fairly to say that the NRQCD factorization for inclusive production of color singlet heavy quark pair does not simply works without any modifications.

The infrared divergence discovered by authors in \cite{NQS:2005,Nayak:2006fm} reads:
\begin{equation}
\label{infd}
\epsilon^{(8\to 1)}(c\bar{c})=-\frac{N_{c}}{4}(N_{c}^{2}-1)
\frac{\alpha_{s}^{2}}{4\epsilon}[1-\frac{1}{f(|\vec{v}|)}\ln(\frac{1+f(|\vec{v}|)}{1-f(|\vec{v}|)})]
\end{equation}
\begin{equation}
f(x)=\frac{2x}{1+x^{2}}
\end{equation}
, where $\vec{v}$ is the relative velocity of the heavy quark in the center of mass frame of the heavy quark pair. It does not rely on undetected states $X$ and can be absorbed into Wilson line that is independent of the states $X$. As the result,  the NRQCD factorization for inclusive production of color singlet heavy quark pair can be remedied by bringing in Wilson lines that are independent of the states $X$ in the NRQCD matrix elements upto NNLO. For higher order, however, the issue is much non-trivial. Especially, gluons exchanged between higher Fock states and undetected states $X$  cause new infrared divergences, which read(\cite{zhou:2015inc}):
\begin{eqnarray}
\label{infdhf}
&&\epsilon^{(8\to 1)}(\text{high Fock states})
\nonumber\\
&=&\frac{\alpha_{s}^{\phantom{s}2}}{16\pi^{2}}
\frac{N_{c}(N_{c}^{2}-1)}{4}
(\frac{1}{\varepsilon})^{2}
[\ln^{2}\frac{(v_{1}\cdot l)^{2}}{(v_{2}\cdot l)^{2}}-2\varepsilon\ln\frac{\Lambda^{2}}{\mu^{2}}
\ln^{2}\frac{(v_{1}\cdot l)^{2}}{(v_{2}\cdot l)^{2}}
\nonumber\\
&&
+\varepsilon(\ln^{2}\frac{(v_{1}\cdot l)^{2}}{(v_{1})^{2}}
-\ln^{2}\frac{(v_{2}\cdot l)^{2}}{(v_{2})^{2}})\ln\frac{(v_{1}\cdot l)^{2}}{(v_{2}\cdot l)^{2}}]
\end{eqnarray}
, with $l$ the direction of the Wilson line in the NRQCD matrix elements. It is hardly to believe that the independence of topologically unfactorized infrared divergences on the direction of the Wilson line  may holds up for arbitrary states containing non-relativistic heavy quark pair.

In practical experiments, heavy quarkonia are reconstructed by their decay products. Thus detected states are these products(like $\mu^{+}\mu^{-}$) in piratical processes. Heavy quarkonia, no mention to the color singlet heavy quark pair, are not detected states in these process. Based on this fact, a scheme is presented in \cite{zhou:2015inc} to cancel out topologically unfactorized infrared divergences in inclusive production of heavy quarkonia. It is shown in \cite{zhou:2015inc} that topologically unfactorized infrared divergences cancel out in the process:
\begin{equation}
A+B\to l^{+}l^{-}(n,p_{H})+X
\end{equation}
, where $A$ and $B$ represent initial particles, $X$ represents undetected final particles, $n$ represents the quantum number of the heavy quarkonium one concerned, the momentum $p_{H}$ of the detected  lepton pair fulfill the condition:
\begin{equation}
p_{H}^{2}\simeq M^{2}
\end{equation}
with $M$ the mass of the heavy quarkonium one concerned. According to this conclusion, we see that effects of transitions between states containing a heavy quark pair(color singlet or octet) should be taken into account for cancellation of topologically unfactorized infrared divergences. Such transitions are caused by exchanges of soft gluons between states containing a heavy quark pair and other undetected states.

In this paper, we present the detailed proof of the cancellation of topologically unfactorized infrared divergences in the process considered in \cite{zhou:2015inc} at leading order in the width of heavy quarkonia.  Physical picture of such cancellation is clear. The lepton pair is produced in short distance(order $1/M$) subprocess.     Thus effects of QCD interactions, which do not affect the detected $l^{+}l^{-}$  pair  after the production of the lepton pair, cancel out once the summation over other states has been made according to unitarity. The interval between the production of the heavy quark pair and the lepton pair is finite unless the relative velocities between intermediate states connecting these two subprocess vanishes. Thus the process is free from topologically unfactorized infrared divergences unless these relative velocities vanish. This is related to the fact that gluons exchanged between states containing the heavy quark pair and other undetected states are pinched in infrared region only if these relative velocities vanish. For the case that these relative velocities do vanish, effects of couplings between intermediate states and infrared gluons cancel out according to the global color symmetry of QCD interactions as the detected lepton pair is color singlet.

The paper is organized as follows. In Sec.\ref{cofac}, we describe the process considered in this paper. In Sec.\ref{psa}, we analyse pinch singular surfaces of gluons exchanged between energetic particles in the process. In Sec.\ref{LPSS}, we take the infrared power counting and explain the characteristics of reduced graphs on leading pinch singular surfaces(LPSS).  In Sec.\ref{cancellation}, we present the proof of cancellation of topologically unfactorized infrared divergences in the process.  We give our conclusion and some discussions in Sec.\ref{conc}.

\section{Inclusive Production of lepton pair Near the Threshold of Heavy Quarkonia}
\label{cofac}

In this section, we describe the process considered here. The process can be written as:
\begin{equation}
A+B\to l^{+}l^{-}(n,p_{H})+X
\end{equation}
, where $A$ and $B$ represent initial particles(hadrons or leptons), $X$ represents undetected final particles, $n$ represents the quantum number of the heavy quarkonium $H$ one concerned, invariant momentum of the lepton pair is constrained to be near the mass shell of the heavy quarkonium $H$, that is,
\begin{equation}
p_{H}^{2}\simeq M^{2}
\end{equation}
with $M$ the mass of $H$. It is possible that the lepton pair is not produced by the decay of $H$ in this process. Although contributions of process with the lepton pair produced according to other mechanism are suppressed by the small width of $H$, effects of transition between $H$ and other possible intermediate states that produce the lepton pair is crucial in the cancellation of topologically unfactorized infrared divergences in the process as pointed out in \cite{zhou:2015inc}.

In the center of mass frame of initial particles, $A$ and $B$ move in opposite directions. They are considered to be collinear to different light like directions in hard collision with corrections of order $m^{2}/s$, where $m$ is the mass scale of the initial particles, $s$ is the square of center of mass energy of initial particles. If the transverse momentum of the final lepton pair is much greater than the mass of $H$, then we have the collinear factorization theorem(\cite{Kang:2011mg,Kang:2011zza,Kang:2014tta,Fleming:2012wy,Fleming:2013qu}):
\begin{eqnarray}
\label{cfmu}
&&\sum_{X}\ud \sigma_{A+B\to l^{+}l^{-}(n,p_{H})+X}
\nonumber\\
&=&\sum_{i,X}\ud \sigma_{A+B\to i+X}\otimes D_{i\to l^{+}l^{-}(n,p_{H})}
\nonumber\\
&&
+
\sum_{\kappa,X} \ud \sigma_{ A+B\to Q\bar{Q}(\kappa)+X}\otimes D_{Q\bar{Q}(\kappa)\to l^{+}l^{-}(n,p_{H})}
\nonumber\\
&&+\mathcal{O}(M^{4}/p_{T}^{4})
\nonumber\\
&&+\ldots
\end{eqnarray}
, where $D_{i\to l^{+}l^{-}(n,p_{H})}$ and $D_{Q\bar{Q}(\kappa)\to l^{+}l^{-}(n,p_{H})}$ represent fragmentation functions for $i$ and $Q\bar{Q}(\kappa)$ to the state $l^{+}l^{-}(n,p_{H})$ under the evolution of QCD and QED interactions, the ellipsis represents contributions of process with the lepton pair produced in the short distance(order $1/p_{T}$) subprocess and possible interference terms, $p_{T}$ represents the transverse component of $p_{H}$.

In the collinear factorization formula (\ref{cfmu}), effects of short distance(order $1/p_{T}$) physics are absorbed into the cross sections of scattering processes between partons, while long distance effects are described by matrix elements between hadrons. Long distance matrix elements are independent of explicit process while the short distance parton cross section can be calculated according to perturbation theory. The collinear factorization formula (\ref{cfmu}) is by itself quite predictive.    For the heavy quarkonium $H$, we have $M^{2}\gg \Lambda_{QCD}^{2}$. The predictive power of the factorization theorem can be improved once effects of physics at the distance of order $1/M$, which can be calculated perturbatively, has been factorized from long distance matrix elements.

To integrate out degrees of freedom with momenta off shell of order $M^{2}$, it is convenient to consider long distance matrix elements in the rest frame of the final detected lepton pair. This is equivalent to consider the problem in the rest frame of the heavy quarkonium $H$ once the lepton pair is produced through the decay of $H$.
In this frame, Non-relativistic QCD(NRQCD)\cite{Caswell:1985ui,Bodwin:1994jh} is the effective theory designed to separate effects of QCD interactions at short distances(order $1/M$) and long distances(order $1/Mv$ or longer with $v$ the relative velocity between the heavy quark pair in $H$). Heavy quarkonia are considered to be non-relativistic bound states in NRQCD.   Short distance effects correspond to the production or annihilation of the heavy quark pair,  which can be determined according to perturbative expansion of QCD coupling constant $\alpha_{s}$. Long distance effects are absorbed into matrix elements of effective operators in NRQCD, which are organized according to expansion series of $v^{2}$. As the result, calculations based on NRQCD are organized according to the expansion of $\alpha_{s}$ and $v^{2}$.

If the NRQCD factorization theorem  proposed in \cite{Bodwin:1994jh} does work, then the long distance  matrix elements are independent of explicit process. For exclusive production of two charmonia in $e^{+}e^{-}$ annihilation and production of a charmonium and a light meson in B-meson decays,  the proofs are established in \cite{Bodwin:2008nf,Bodwin:2010fi}.
The NRQCD factorization  theorem for inclusive production of heavy quarkoia, however,  can be violated by soft gluons exchanged between the  detected heavy quarkonia states and other undetected energetic particles as pointed out in \cite{NQS:2005,Nayak:2006fm}. For the process considered here, soft gluons exchanged between energetic particles may also violate the NRQCD factorization theorem. Especially, the cancellation of topologically unfactorized infrared divergences is crucial for NRQCD factorization theorem hold up in this process.  We thus consider pinch singular surfaces of soft gluons exchanged between energetic particles in next section.

\section{Pinch Singularities Analysis of the Process}
\label{psa}

In this section, we analyze pinch singular surfaces of diagrams that contribute to the process. Singular points of integral function of these diagrams can cause mass singularities only if the integral paths are pinched at these points. That is, these singular points are pinch singular points. In other case, one can deform the integral path to avoid the singular points without change the Feynman integral.

\begin{figure}
\begin{center}
\includegraphics[width=0.4\textwidth]{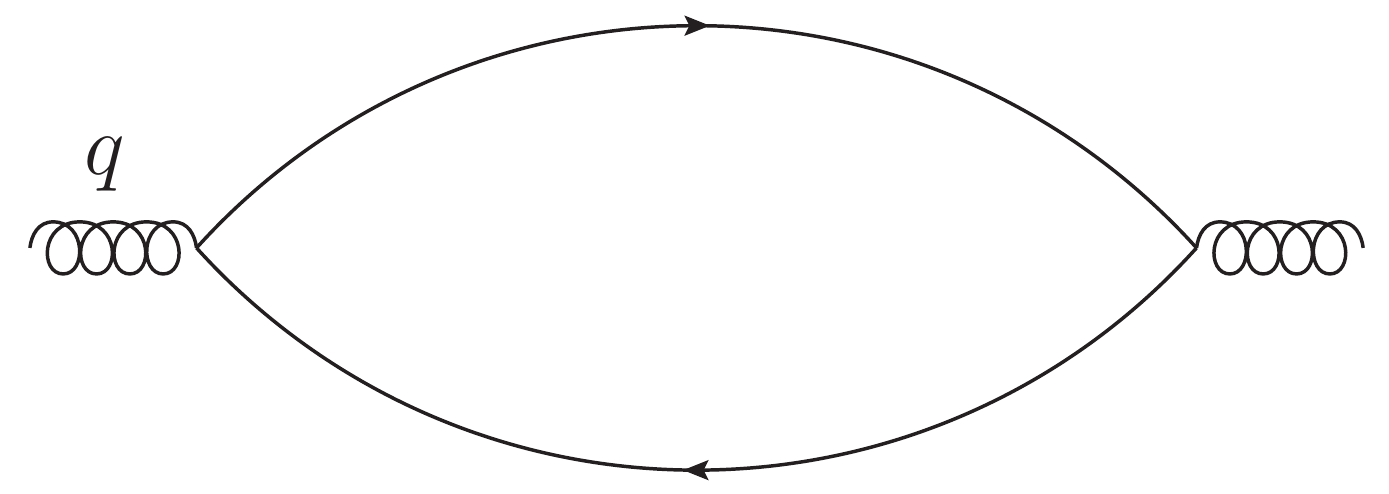}
\end{center}
\caption{Vacuum polarization diagram}
\label{fig:vapolz}
\end{figure}

We consider reduced graphs, which is constructed from complete diagrams by simply contracting all off shell internal lines.   According to the pinch singularities analysis in \cite{CN:1965,S:1993}, reduced graphs on pinch singular surfaces correspond to physical process of classical particles. For example, one consider the vacuum polarization diagram shown in Fig.\ref{fig:vapolz}. If $q^{2}$ is smaller than the threshold of fermions in the diagram, then these fermions can not be both on shell and the diagram is free from mass singularities. In the case that $q^{2}$ is larger than the threshold of these fermions, they move in different directions and world lines of them intersect with each other at two distinct points in space time. This is impossible for classical trajectory of these fermions. Thus the diagram is free from mass singularities in this case. For $q^{2}$ equal to
the threshold of these fermions, they move in the same direction and two world lines coincide with each other. As the result, the singular points of internal lines in diagram shown in Fig.\ref{fig:vapolz} do correspond to the pinch singular points in this case.

\begin{figure*}
\begin{tabular}{c@{\hspace*{10mm}}c@{\hspace*{10mm}}c}
\includegraphics[scale=0.3]{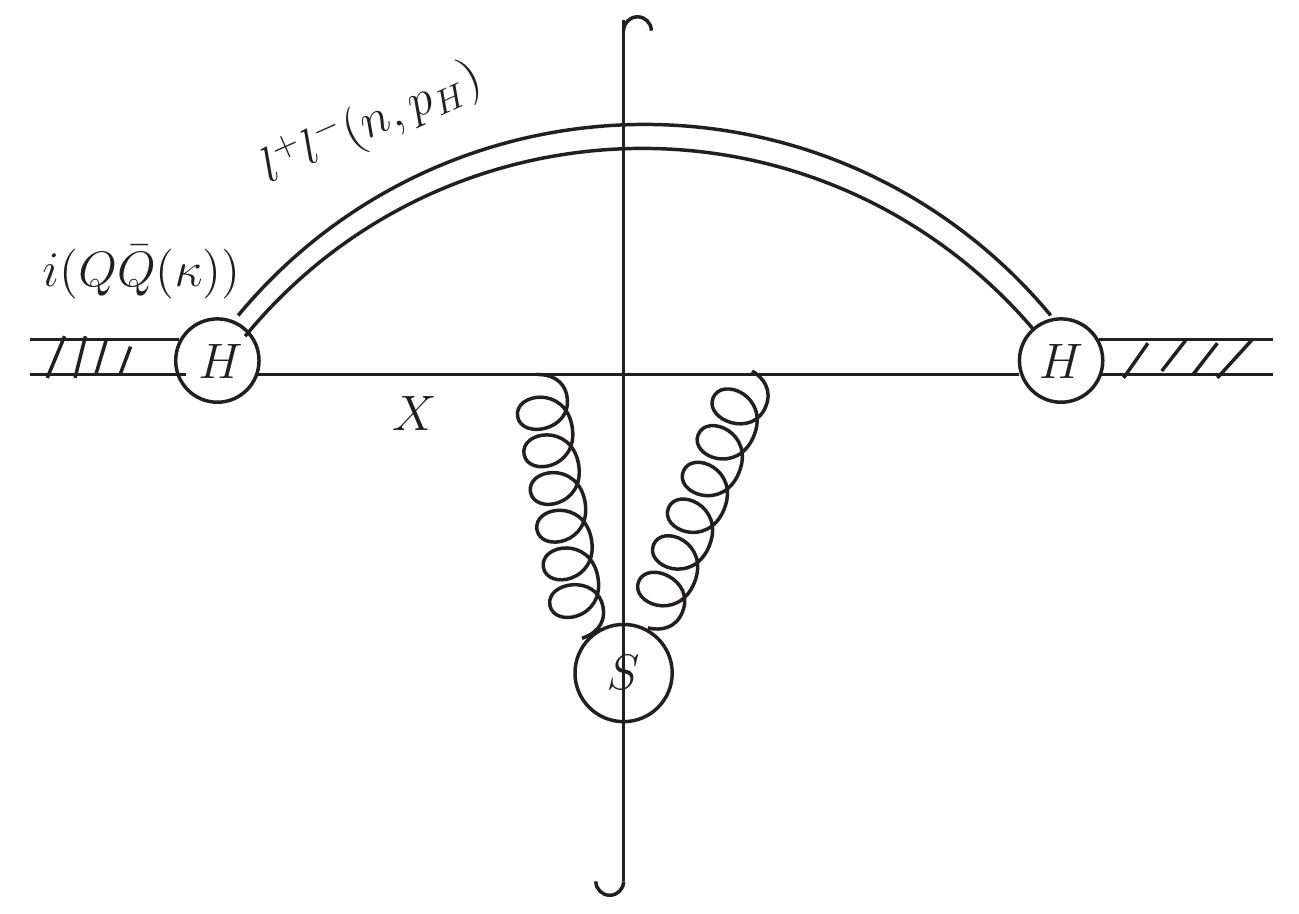}
&
\includegraphics[scale=0.3]{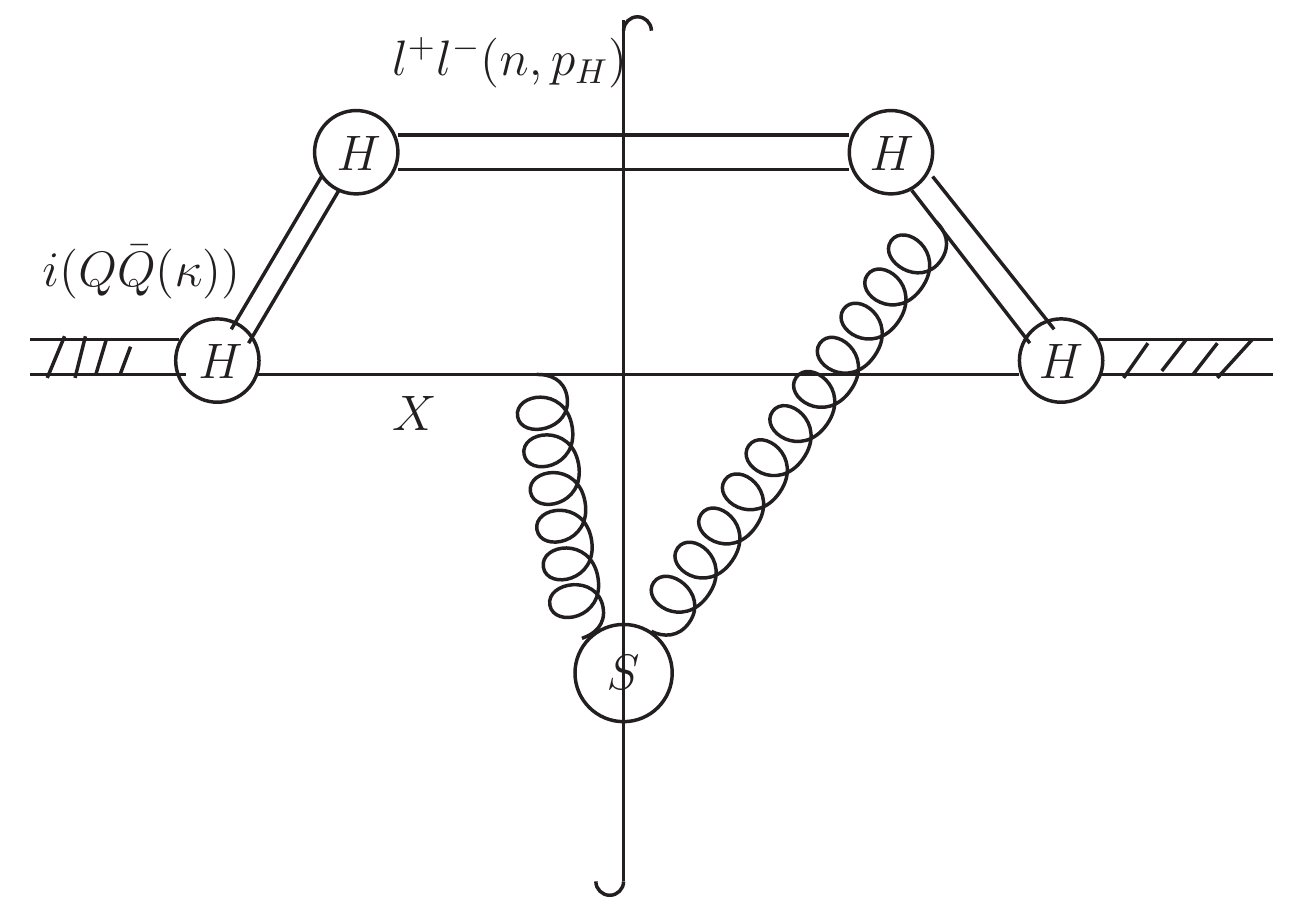}
&
\includegraphics[scale=0.3]{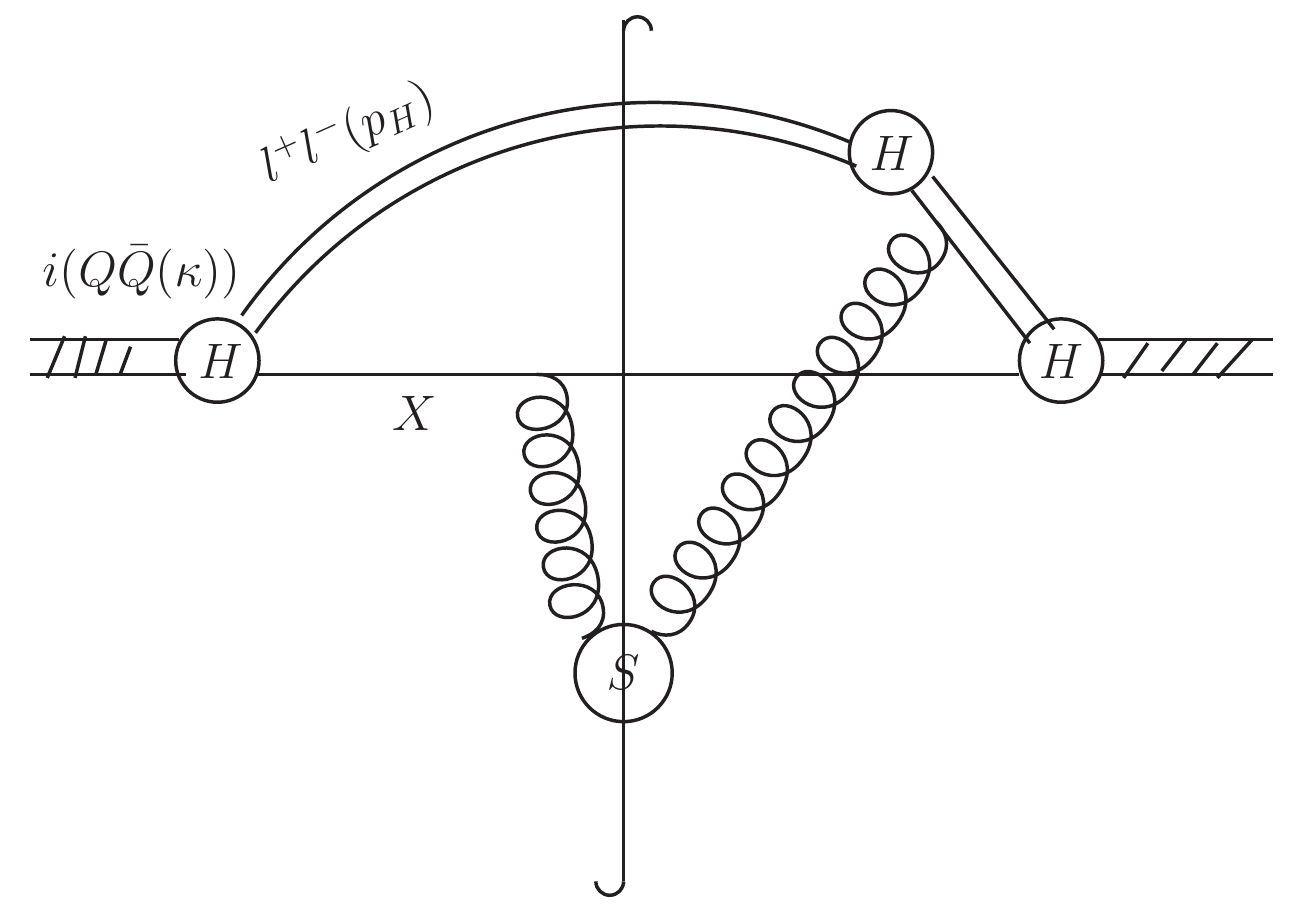}
\\
(a)&(b)&(c)
\end{tabular}
\caption{(a) Example of direct production of $l^{+}l^{-}$ pair in fragmentation functions $D_{i\to l^{+}l^{-}(n,p_{H})}$ and $D_{Q\bar{Q}(\kappa)\to l^{+}l^{-}(n,p_{H})}$; (b)Example of indirect production of $l^{+}l^{-}$ pair in fragmentation functions $D_{i\to l^{+}l^{-}(n,p_{H})}$ and $D_{Q\bar{Q}(\kappa)\to l^{+}l^{-}(n,p_{H})}$; (c) Example of interference terms between the direct and indirect production  of $l^{+}l^{-}$ pair.}
\label{incmu}
\end{figure*}

For fragmentation functions $D_{i\to l^{+}l^{-}(n,p_{H})}$ and $D_{Q\bar{Q}(\kappa)\to l^{+}l^{-}(n,p_{H})}$, the lepton pair is produced by:(1) direct fragmentation of the parton $i$ or $Q\bar{Q}(\kappa)$ in short distance(order $1/M$); (2)short distance(order $1/M$) electromagnetic scattering processes between nearly on shell intermediate  states produced by the fragmentation of the parton $i$ or $Q\bar{Q}(\kappa)$; (3) interference terms between the direct and indirect production of the lepton pair. Examples of these cases are shown in Fig.\ref{incmu}.

Direct production processes of lepton pair are free from topologically unfactorized QCD infrared divergences as the lepton pair decouple from gluons. In addition, infrared divergences caused by gluons exchanged between the undetected particles $X$ cancel out once the summation over these states has been made.

\begin{figure*}
\begin{tabular}{c@{\hspace*{10mm}}c}
\includegraphics[scale=0.3]{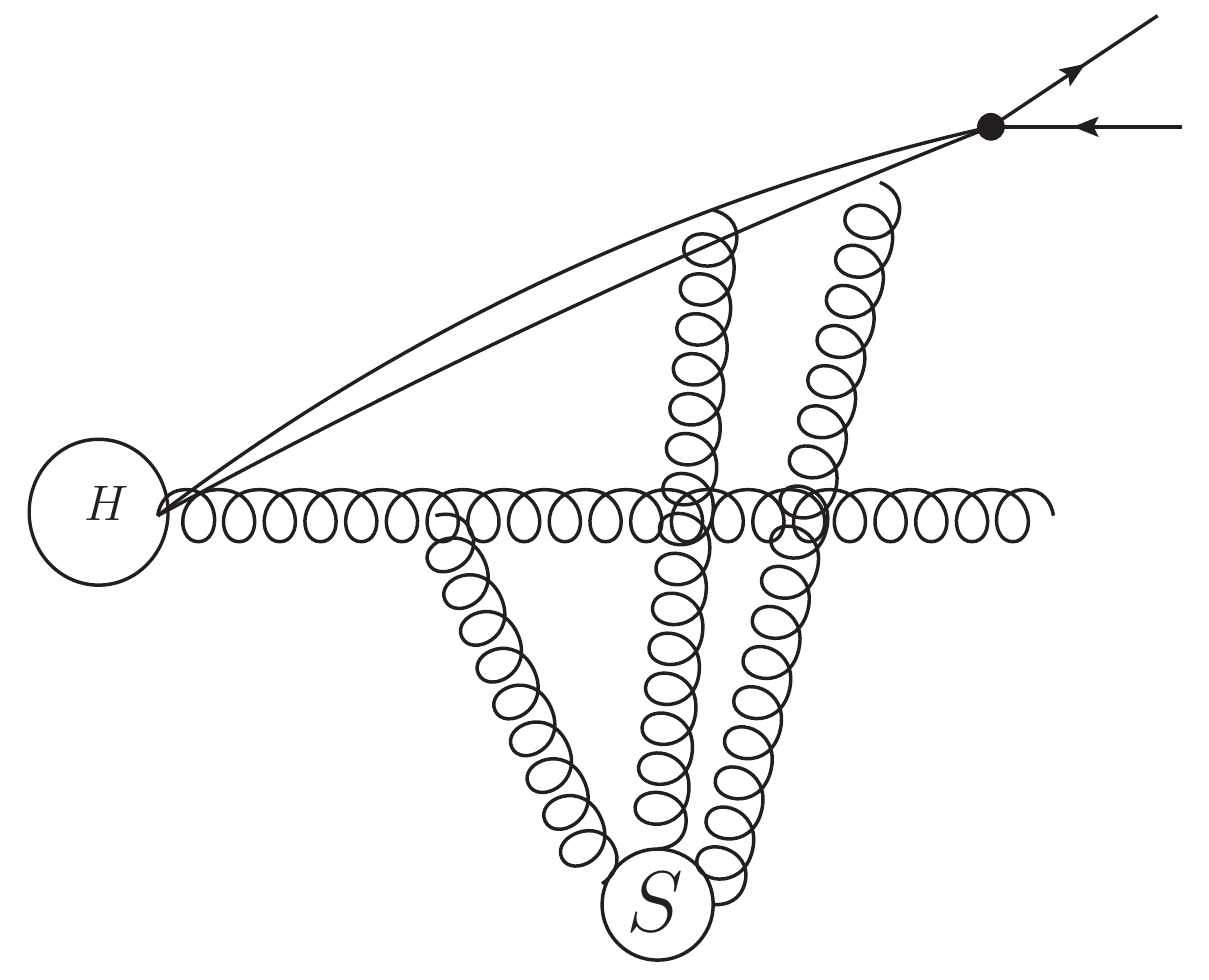}
&
\includegraphics[scale=0.3]{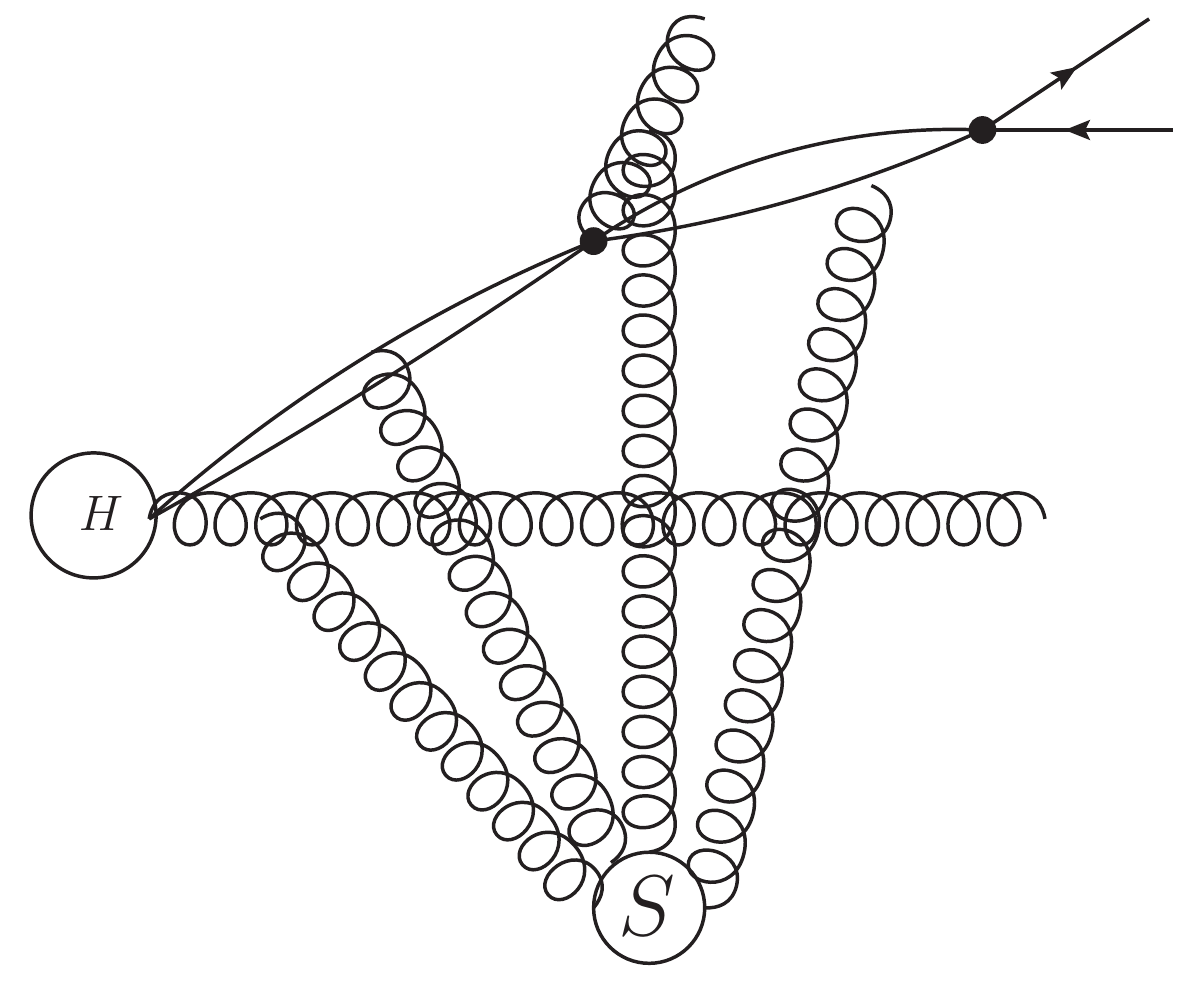}
\\
(a)&(b)
\end{tabular}
\caption{(a) Example of reduced graphs with two vertexes for the indirect production of the lepton pair; (b)Example of reduced graphs with three vertexes for the indirect production of the lepton pair}
\label{indred}
\end{figure*}

For indirect production process of lepton pair, there are at least  two hard subdiagrams corresponding to fragmentation of the parton $i$ or $Q\bar{Q}(\kappa)$ into nearly on shell intermediate states and production of the lepton pair. Both of them are nearly local with uncertainty of order $1/M$. One get at least two vertexes in the reduced graph of these processes after contracting off shell internal lines. Examples of these reduced graphs are shown in Fig.\ref{indred}.

\begin{figure}
\begin{center}
\includegraphics[width=0.4\textwidth]{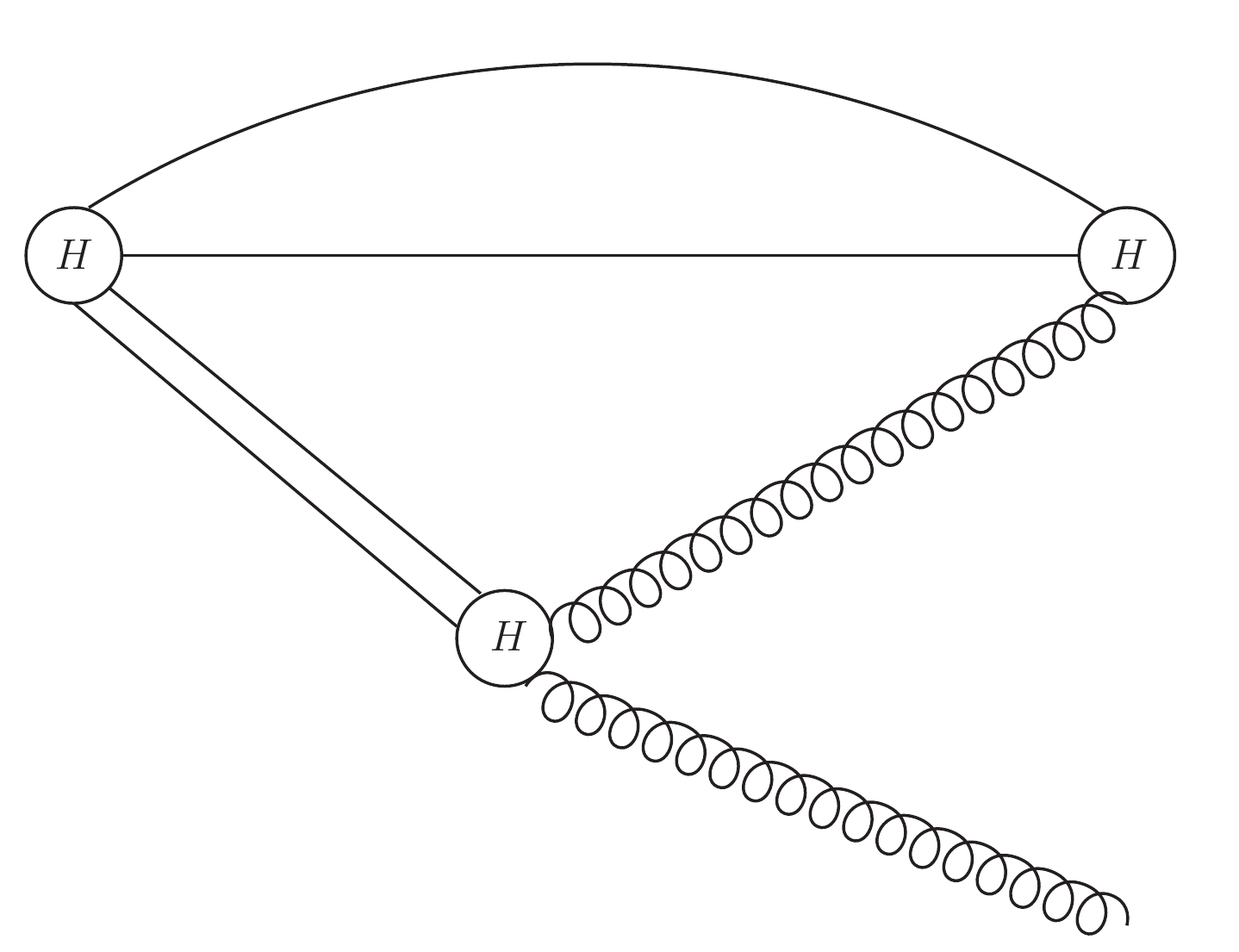}
\end{center}
\caption{Example of reduced graphs with hard rescatterning  subgraph between gluons and heavy quarks. All the solid lines represent heavy quarks.}
\label{fig:hscattering}
\end{figure}

Be different from processes that do not involve massive particles(\cite{S:1978,S:1993}). There are massive particles in the process considered here. Velocities of classical massive particles are smaller than $1$. It is thus possible that light quarks or gluons may catch up heavy quarks in reduced graphs corresponding to pinch singularities. For example, one may consider the diagram shown in fig.\ref{fig:hscattering}.
In Fig.\ref{fig:hscattering}, velocities of gluons are greater than those of heavy quarks. It is thus possible that the rescattering between gluons and heavy quarks after the production of gluons with all heavy quarks in fig.\ref{fig:hscattering} produced in the same point.

If we do not consider the transition between different particles, then world lines of classical energetic particles intersect twice only if they coincide with each other. As the result, relative velocities between energetic particles that connect two hard subgraphs  vanish  for reduced graphs corresponding to pinch singular surfaces. This is the same with those analyzes in  \cite{S:1978,S:1993}.

According to above analysis, we conclude that: (1) diagrams with the lepton pair produced by fragmentation of the parton   $i$ or $Q\bar{Q}(\kappa)$ in short distance(order $1/M$) do not bother us as leptons decouple from gluons;(2) diagrams with the lepton pair produced in  short distance(order $1/M$) electromagnetic scattering processes between nearly on shell intermediate  states do not bother us unless relative velocities between on shell intermediate particles that connect two hard subgraphs vanish; (3) there can be several hard subprocesses that connected by nearly on shell particles as velocities of heavy quark and massless particles are different from each other.

\section{Leading Pinch Singular Surfaces For the Process}
\label{LPSS}

In this section, we study reduced graphs that corresponding to pinch singular surfaces for the process. We take the similar analyzes  as in \cite{S:1978,S:1993} to search for reduced graphs corresponding to leading pinch singular surfaces.

For processes that do not involve massive quarks, reduced graphs corresponding to leading pinch singular surfaces fulfill conditions\cite{S:1978,S:1993}:
\begin{enumerate}
\item {For each jet, there is only one physical particle(light quarks or transverse polarized gluons) that connect to hard subgraphs.}
\item {There may be arbitrary number of scalar polarized collinear gluons that connect to hard subgraphs.}
\item {Collinear jets may connect to the soft subgraphs through soft gluons.}
\item {Soft particles do not connect to hard subgraphs.}
\item {For each vertex at which soft gluons couple to collinear particles, there is only one soft gluon line.}
\end{enumerate}
, where hard subgraphs are made up of off shell internal lines, each jet are made up of on shell internal lines with infinite rapidity that collinear to the same direction,  soft graphs are made up of on shell internal lines with $0$-momenta.

For the process considered here, massive heavy quarks may annihilate into lighter particles. To take the similar infrared power counting as
in \cite{S:1978,S:1993}, we should choose suitable normal and intrinsic variables. Normal variables are integral variables that vanish on 
pinch singular surfaces. Intrinsic variables are the remain integral variables.

For soft massless particles, all momenta components are normal variables. The infrared power counting of these variable read:
\begin{equation}
q^{\mu}\sim \lambda
\end{equation}
with $\lambda$ an infinitesimal quantity.

For a collinear jet with momentum $P^{\mu}$,
the invariant momentum square $P^{2}\to \lambda $ is a normal variable. We denote the direction of $P^{\mu}$ as $n^{\mu}$:
\begin{equation}
n^{\mu}=\frac{1}{\sqrt{2}}(1,\vec{n}),\quad |\vec{n}|^{2}=1
\end{equation}
For a loop with momentum $l$ in the jet, the component $\bar{n}\cdot l$ is intrinsic variable, other components are normal variables, where
\begin{equation}
\bar{n}^{\mu}=\frac{1}{\sqrt{2}}(1,-\vec{n})
\end{equation}
. Infrared power counting for these variables read:
\begin{equation}
(\bar{n}\cdot l, n\cdot l, |\vec{l}_{n\perp}|)\sim (1,\lambda,\lambda^{\frac{1}{2}})
\end{equation}
, with $\vec{n}\cdot \vec{p}_{n\perp}=0$.

For a nearly on shell heavy quark with momentum $p^{\mu}$, the invariant momentum square $p^{2}\sim\lambda $ is a normal variable. Be
different from massless particles, an on shell heavy quark can not resolve to more on shell heavy quarks. We thus do not consider the jet made up of heavy quarks.

We consider a reduced graph with $J$-collinear internal lines, $S$-soft internal lines and $H$-heavy quark propagators.  We denote the number of jet and soft loops in the diagrams as $L^{s}$ and $L^{j}$. The number of collinear jets and heavy quarks that produced in hard subgraphs are denoted as $K^{j}$ and $K^{h}$. Infrared power counting of such diagram reads $\lambda^{p}$, where:
\begin{equation}
p=(4L^{s}+2L^{j}+K^{j}+K^{h})-(2S+J+H)+N
\end{equation}
with $N$ contributions of numerator factors. We further have:
\begin{equation}
p=(2L^{j}+K^{j}+K^{h})-(J+H)+N^{j}+b^{(1j)}+\frac{3}{2}f^{j}+b^{(1h)}+\frac{3}{2}f^{h}
\end{equation}
, where $N^{j}$ represents contribution of numerator factor of collinear jets, $b^{(1j)}$ and $f^{j}$ are
the number of soft gluons and soft light fermions attached  to jet lines, $b^{(1h)}$ and $f^{h}$ are the  number of soft gluons and soft light
fermions attached  to heavy quarks.

For heavy quark propagators, we have:
\begin{equation}
H-K^{h}=V^{h}
\end{equation}
, where $V^{h}$ is the number of vertices at which couplings between heavy quarks and soft particles occur. For each such vertex, there may be several soft gluons or fermions attach to it. We denote the number of vertices at which $f$-soft fermions and $b$-gluons attach to the heavy quark propagators as $V_{f,b}^{h}$. We have:
\begin{eqnarray}
V^{h}&=&\sum_{f,b}(V_{f,b}^{h})
\nonumber\\
f^{h}&=&\sum_{f,b}(f V_{f,b}^{h})
\nonumber\\
b^{1h}&=&\sum_{f,b}(b V_{f,b}^{h})
\end{eqnarray}
\begin{eqnarray}
p&=&(2L^{j}+K^{j})-J+N^{j}+b^{(1j)}+\frac{3}{2}f^{j}+\sum_{f,b}(\frac{3}{2}f+b-1)V_{f,b}^{h}
\nonumber\\
&\ge& (2L^{j}+K^{j})-J+N^{j}+b^{(1j)}+\frac{3}{2}f^{j}
\end{eqnarray}
The equation holds only if:
\begin{equation}
V_{f,b}^{h}=0 \quad \text{for $b\ne 0$ or $b>1$}
\end{equation}
. According to this and results in \cite{S:1978,S:1993}, we see that reduced graphs corresponding to leading pinch singular surfaces fulfill conditions:
\begin{enumerate}
\item {For each jet, there is only one physical particle(light quarks or transverse polarized gluon) that connect to hard subgraphs.}
\item {There may be arbitrary number of scalar polarized collinear gluons that connect to hard subgraphs.}
\item {Collinear jets and heavy quark propagators may connect to the soft subgraphs through soft gluons.}
\item {Soft particles do not connect to hard subgraphs.}
\item {For each vertex at which soft gluons attach to collinear jets or heavy quarks, there is only one soft gluon line.}
\end{enumerate}

We see that, despite of existence of heavy quarks, features of  reduced graphs corresponding to leading pinch singular surfaces for process considered here are quite similar to those in discussed \cite{S:1978,S:1993}. The differences is that there may be several hard subgraphs connected to each other through on shell lines in process considered here compare to those in \cite{S:1978,S:1985,S:1993}.

\section{Cancellation of Topologically Unfactorized Infrared Divergence in the Process}
\label{cancellation}

\begin{figure}
\begin{tabular}{c@{\hspace*{10mm}}c}
\includegraphics[scale=0.3]{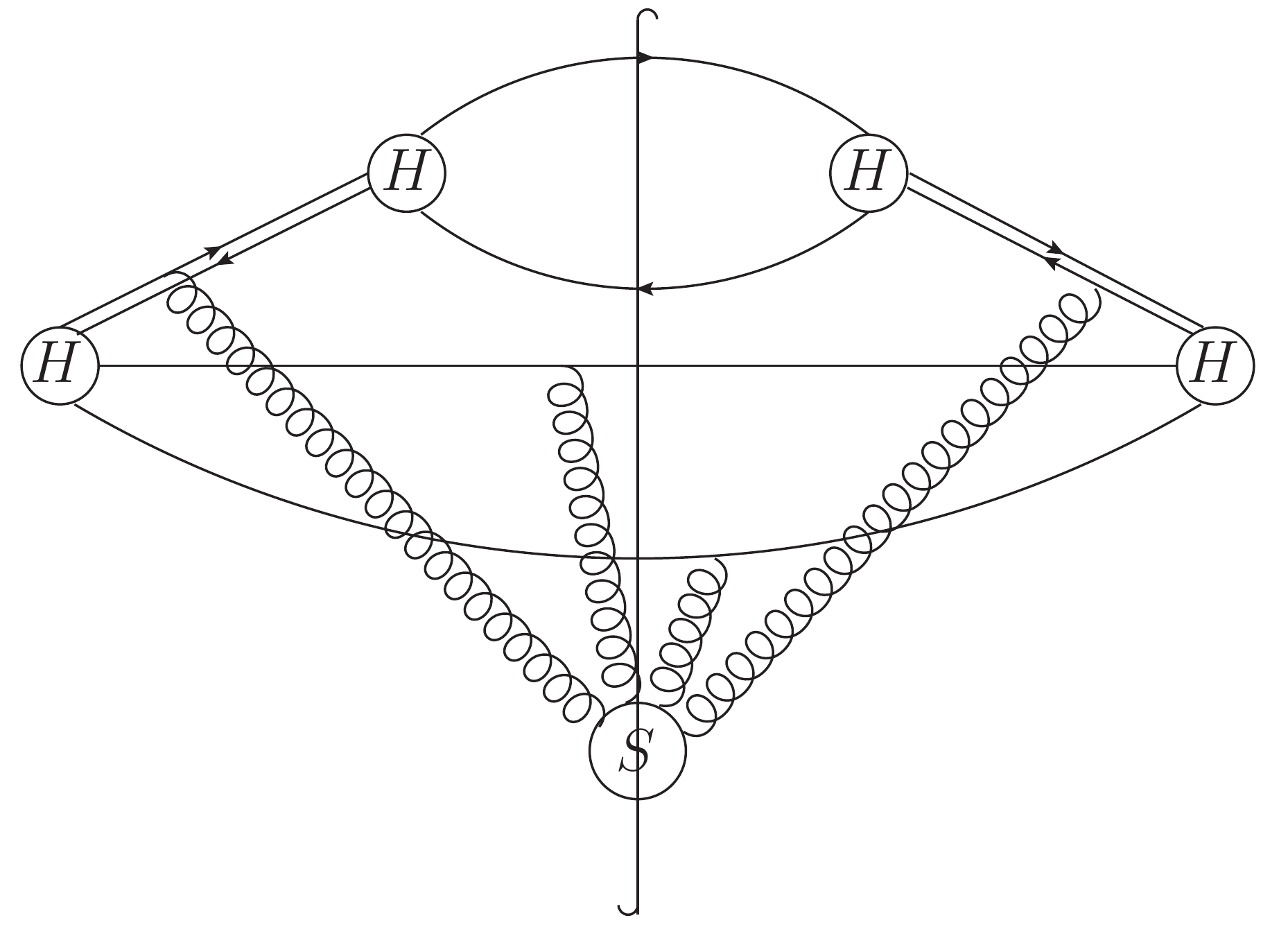}
&
\includegraphics[scale=0.3]{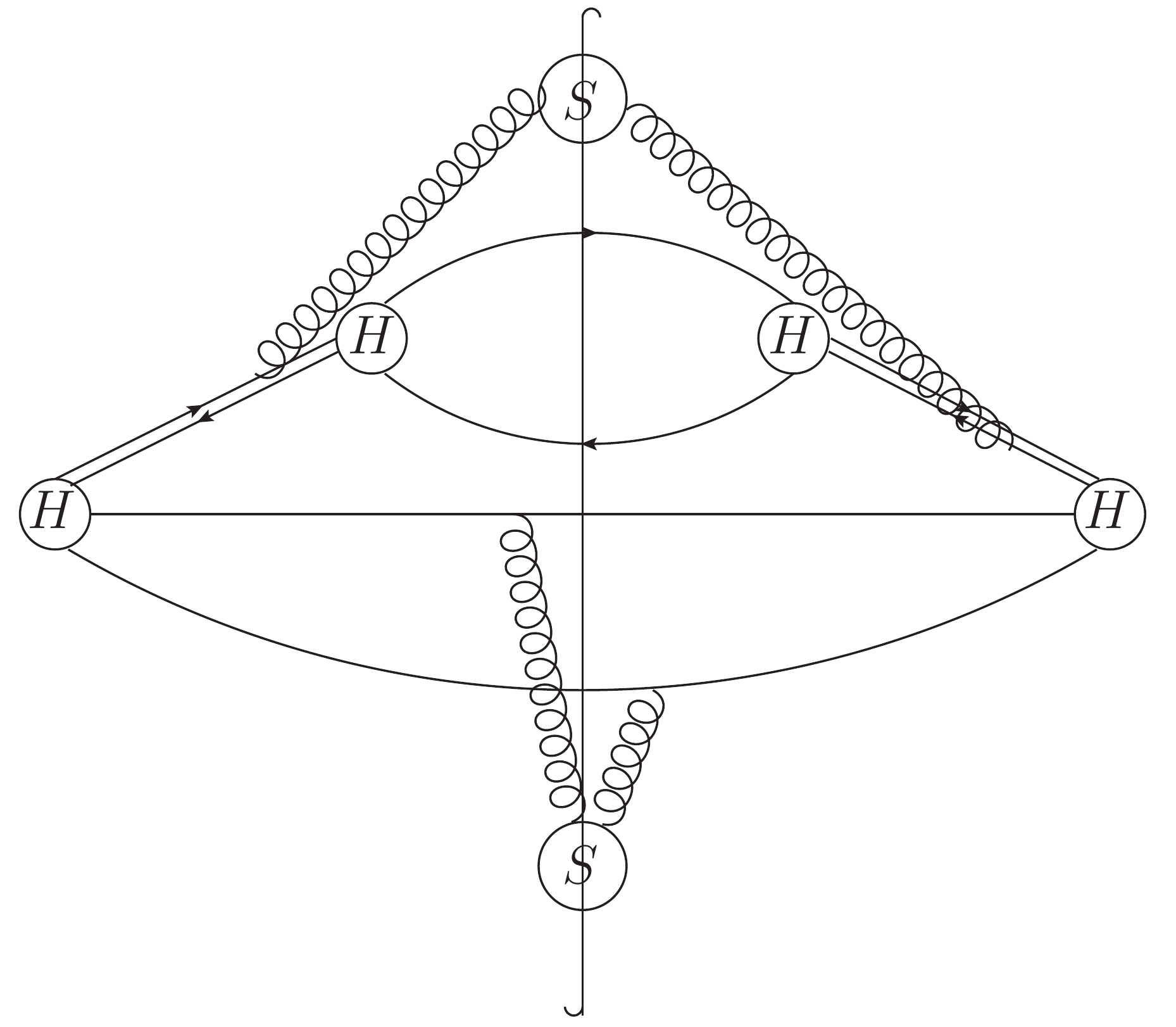}
\\
(a)&(b)
\end{tabular}
\caption{Reduced graphs corresponding to leading pinch singular surfaces in the process. Cut solid lines with arrows represent the detected lepton pair. Cut solid lines without arrows represent possible undetected energetic particles. Uncut solid lines with arrow represent heavy quark pair. Gluons that connect to the soft subgraphs  are all infrared. Scalar polarized collinear gluons are not shown explicitly.}
\label{fig:irlo}
\end{figure}

In this section, we present the proof of cancellation of  topologically unfactorized infrared divergence in the process at leading order of $\Gamma_{H}$. According to the result in last section, reduced graphs corresponding to leading pinch singular surfaces take the topology shown in Fig.\ref{fig:irlo}.
We thus consider reduced graphs of this type. In these diagrams, relative momentum between the heavy quark pair which annihilate into the detected $\mu^{+}\mu^{-}$ vanish unless there are not infrared gluons connecting to the heavy quark pair. This is because that such reduced graphs should be in accordance with trajectories of classical particles.(\cite{CN:1965,S:1993}) World lines of two particles intersect twice or more only if they coincide with each other.  As the result, relative momentum between the heavy quark pair in Fig.\ref{fig:irlo} vanish.

It is necessary to point it out that heavy quarks are not stable particles as they may decay to other states. Such effects are caused by the weak interaction, which are neglected here. In addition, the heavy quark pair in Fig.\ref{fig:irlo} may be produced by annihilation of heavier quarks. For example, charm quarks can be produced by annihilation of $b\bar{b}$ pairs. Contributions of diagrams with such annihilations are suppressed. We thus do not consider these diagrams.

The second diagram in Fig.\ref{fig:irlo} is topologically factorized, which do not disturb us. For diagrams that take the topology as the first diagram in Fig.\ref{fig:irlo},  relative momentum between the heavy quark pair which annihilate into the detected lepton pair vanish unless there are not infrared gluons connecting to the heavy quark pair. In this case,  differences between couplings of infrared gluons to the heavy quark and heavy anti-quark can only be produced by color structures of the heavy quark and heavy anti-quark. If the final heavy quark pair is color singlet as in Fig.\ref{fig:irlo}, one may expect the cancellation of such couplings. This is what we will prove in the following texts.

\subsection{Grammer-Yennie Approximation in Two Particle Irreducible Graphs}
\label{GY:2PI}

The Grammer-Yennie approximation(\cite{GY:1973}) is of great importance in dealing with infrared divergences. We consider the coupling between a infrared gluon with momentum $q$ and an on shell heavy quark with momentum $k$. The Grammer-Yennie approximation is made in such coupling:
\begin{equation}
(k\pm q)^{2}-m^{2}\simeq \pm 2k\cdot q +k^{2}-m^{2},\quad A^{\mu}(q)=\frac{k\cdot A(q) k^{\mu}}{m^{2}}
\end{equation}
, where $m$ is the mass of the heavy quark.
For process considered here, the heavy quark pair in Fig.\ref{fig:irlo} is at rest in the rest frame of the detected lepton pair. In addition, relative momentum between the heavy quark pair vanish unless infrared gluons do not connect to the heavy quark pair.  We can thus write the  Grammer-Yennie approximation as:
\begin{equation}
(k\pm q)^{2}-m^{2}\simeq \pm 2k^{0}q^{0}+k^{2}-m^{2},\quad A^{\mu}(q)=A^{0}(q)\delta^{\mu 0}
\end{equation}
.

The approximation, however, can be violated by Coulomb gluons:
\begin{equation}
|q^{0}|\ll |\vec{q}|
\end{equation}
. Momenta of gluons exchanged between heavy quarks and other undetected energetic particles are not pinched in such region. We can deform the integral path to avoid the Coulomb region for such gluons. For gluons exchanged between the heavy quark pair, the momenta integral is pinched in the Coulomb region once relative momentum between the heavy quark pair vanish. Although these gluons do not cause topologically unfactorized infrared divergences, they do violet the Grammer-Yennie approximation.

\begin{figure}
\begin{tabular}{c@{\hspace*{10mm}}c}
\includegraphics[scale=0.3]{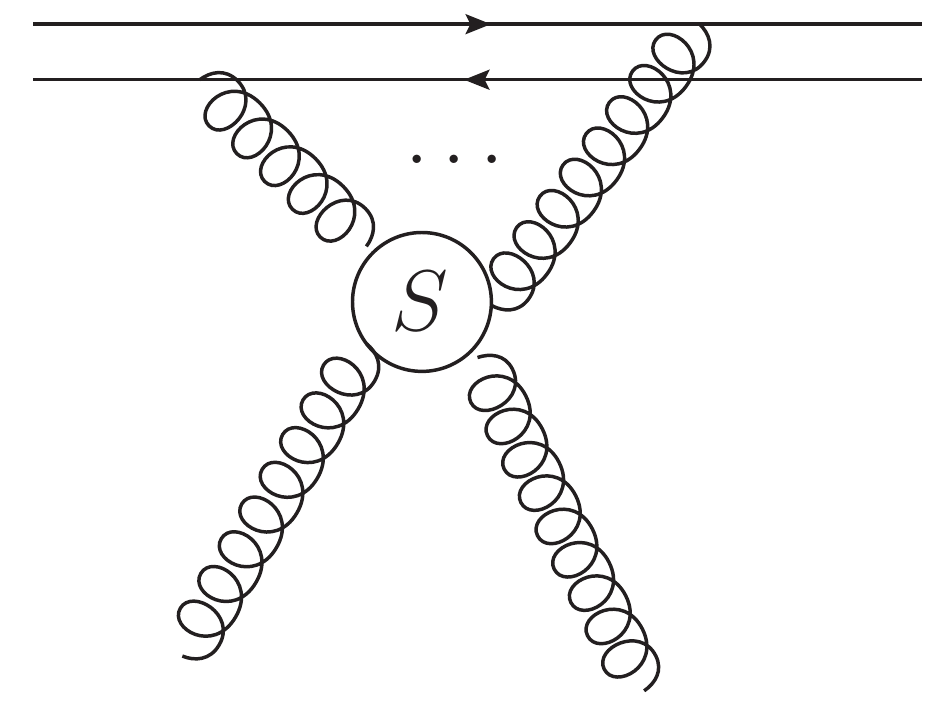}
&
\includegraphics[scale=0.3]{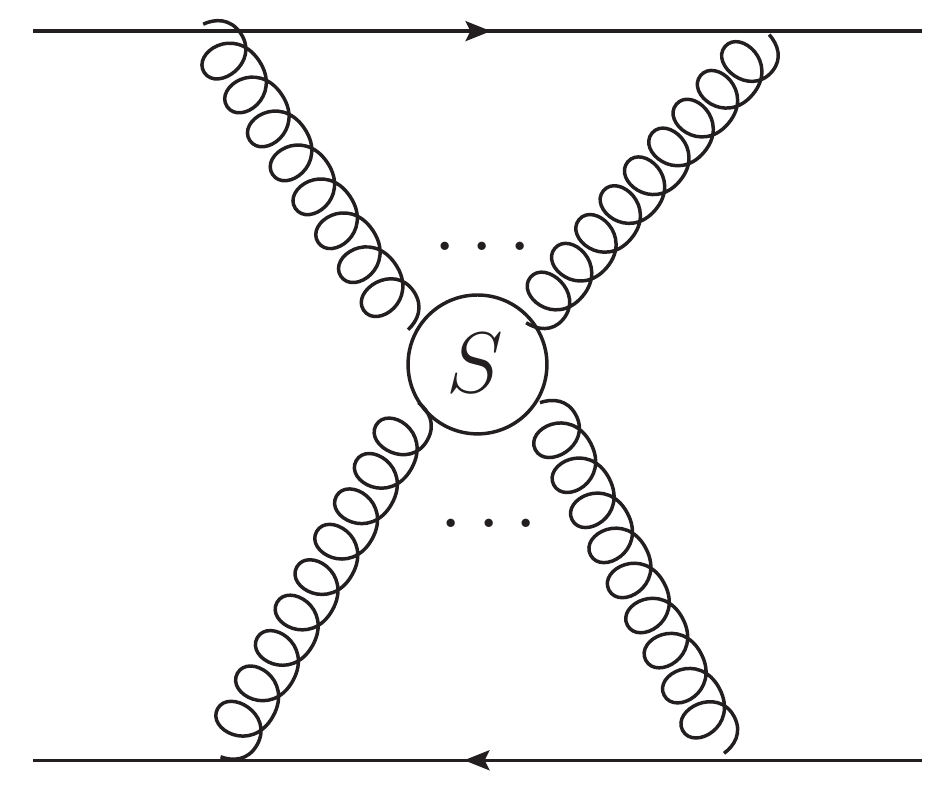}
\\
(a)&(b)
\end{tabular}
\caption{(a) Example of two particle irreducible graphs with some infrared gluons exchanged between heavy quarks and other undetected energetic particles;(b)Example of two particle irreducible graphs with all infrared gluons exchanged between the heavy quark pair}
\label{fig:2PI}
\end{figure}

To conquer the difficulties caused by Coulomb gluons, let us consider the two particle irreducible   diagrams at first. Examples of these diagrams are shown in Fig.\ref{fig:2PI}.
Diagrams with all infrared gluons exchanged between the heavy quark pair, like the second diagram in Fig.\ref{fig:2PI}, do not affect the factorization and can be absorbed into long distance matrix-elements between bound states of the heavy quark pair.

For  two particle irreducible graphs with some infrared gluons exchanged between heavy quarks and other undetected energetic particles, like the first diagram in Fig.\ref{fig:2PI}, we can deform the integral path so that momenta of  gluons exchanged between heavy quarks and other undetected energetic particles are far from the Coulomb region. We then make the Grammer-Yennie approximation in couplings between these gluons and heavy quarks. We denote the momenta scale of gluons coupled to heavy quarks as $|\vec{q}|$. After the deformation, the heavy quark internal lines are off shell of order $m|\vec{q}|$. For coupling between a infrared gluon with momentum $l$ and a heavy quark with momentum $k$, we have:
\begin{equation}
|k^{2}-m^{2}|\sim m |\vec{q}|\ll |\vec{q}|^{2}\sim |l^{2}|
\end{equation}
. We can then take the approximation:
\begin{equation}
(k\pm l)^{2}\simeq \pm 2k\cdot l+k^{2}-m^{2}
\end{equation}
. That is to see, we can take the Grammer-Yennie  approximation in couplings between infrared gluons and heavy quarks in these diagrams.

\subsection{Cancellation of Topologically Unfactorized Infrared Divergences in Two Particle Irreducible Graphs}
\label{cancellation:2PI }

In this subsection, we prove the cancellation  of topologically unfactorized infrared divergences in two particle irreducible graphs like the first diagram in Fig.\ref{fig:2PI}.
After the Grammer-Yennie approximations, contributions of couplings between infrared gluons and heavy quarks can be absorbed into the Wilson lines:
\begin{equation}
Y(A^{\mu})=P\exp(ig\int_{0}^{\infty}\ud s A^{0}(s,\vec{0}))
\end{equation}
. The part of diagrams that depend on infrared gluons can be written as:
\begin{equation}
S\equiv \big<T(\mathcal{O}_{ij}(A)M_{ij}(A^{\mu}))\big>
\end{equation}
, where $\mathcal{O}_{ij}(A)$ is the effective operator produced by the other end of infrared gluons exchanged between heavy quarks and other undetected energetic particles, $i$ and $j$ is the color indices in the fundamental representation, the operator $M_{ij}(A^{\mu})$ is defined as:
\begin{equation}
M_{ij}(Y_{0}(A))=\frac{\int\mathcal{D}[\psi] \mathcal{D}[\bar(\psi)]\mathcal{D}[B^{\mu}]Y_{ik}(A^{\mu}+B^{\mu})Y^{\dag}_{kj}(A^{\mu}+B^{\mu}) e^{i\int\ud^{4}(x)\mathcal{L}(\psi,\bar{\psi},B^{\mu})(x)}}{\int\mathcal{D}[\psi] \mathcal{D}[\bar(\psi)]\mathcal{D}[B^{\mu}] e^{i\int\ud^{4}(x)\mathcal{L}(\psi,\bar{\psi},B^{\mu})(x)}}
\end{equation}
with $\mathcal{L}(\psi,\bar{\psi},B^{\mu})(x)$ the Lagrangian density of QCD:
 \begin{eqnarray}
\mathcal{L}(\psi,\bar{\psi},B^{\mu})(x)&=&
\bar{\psi}(i\not\!\partial-m)\psi(z)
+\frac{1}{2g^{2}} tr ([\partial^{\mu}-igB^{\mu},\partial^{\nu}-igB^{\nu}])^{2}(x)
\end{eqnarray}
. It is necessary to point it out that interactions between fields $A^{\mu}(x)$ and $B^{\mu}(x)$ are not included in above formula. In fact, gluon fields that couple to  fields $A^{\mu}(x)$ should be described by fields $A^{\mu}(x)$ not $B^{\mu}(x)$(\cite{zhou:2015inc}).

According to the unitarity of the Wilson line $Y(A^{\mu})$, we have:
\begin{eqnarray}
M_{ij}(Y_{0}(A))&=&\frac{\int\mathcal{D}[\psi] \mathcal{D}[\bar(\psi)]\mathcal{D}[B^{\mu}]\delta_{ij} e^{i\int\ud^{4}(x)\mathcal{L}(\psi,\bar{\psi},B^{\mu})(x)}}{\int\mathcal{D}[\psi] \mathcal{D}[\bar(\psi)]\mathcal{D}[B^{\mu}] e^{i\int\ud^{4}(x)\mathcal{L}(\psi,\bar{\psi},B^{\mu})(x)}}
\nonumber\\
&=&\delta_{ij}
\end{eqnarray}
We thus conclude that topologically unfactorized infrared divergences in two particle irreducible graphs like the first diagram in Fig.\ref{fig:2PI} cancel out.

\subsection{Cancellation of Topologically Unfactorized Infrared Divergences in Two Particle Reducible Graphs}
\label{cancellation:2PR}

\begin{figure}
\begin{tabular}{c@{\hspace*{10mm}}c}
\includegraphics[scale=0.3]{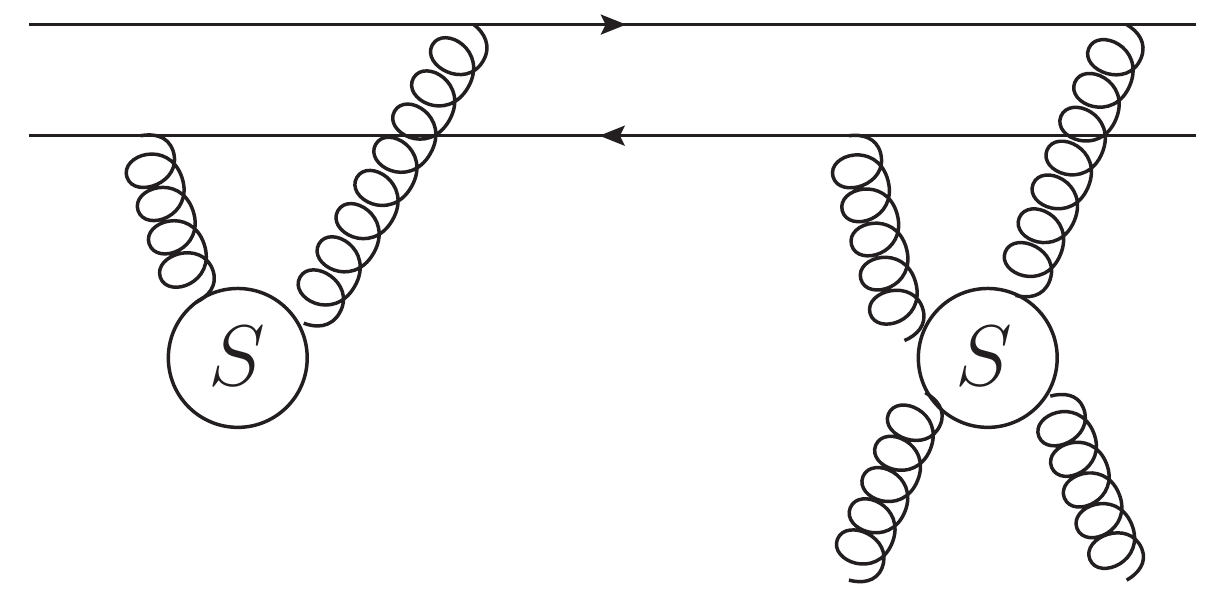}
&
\includegraphics[scale=0.3]{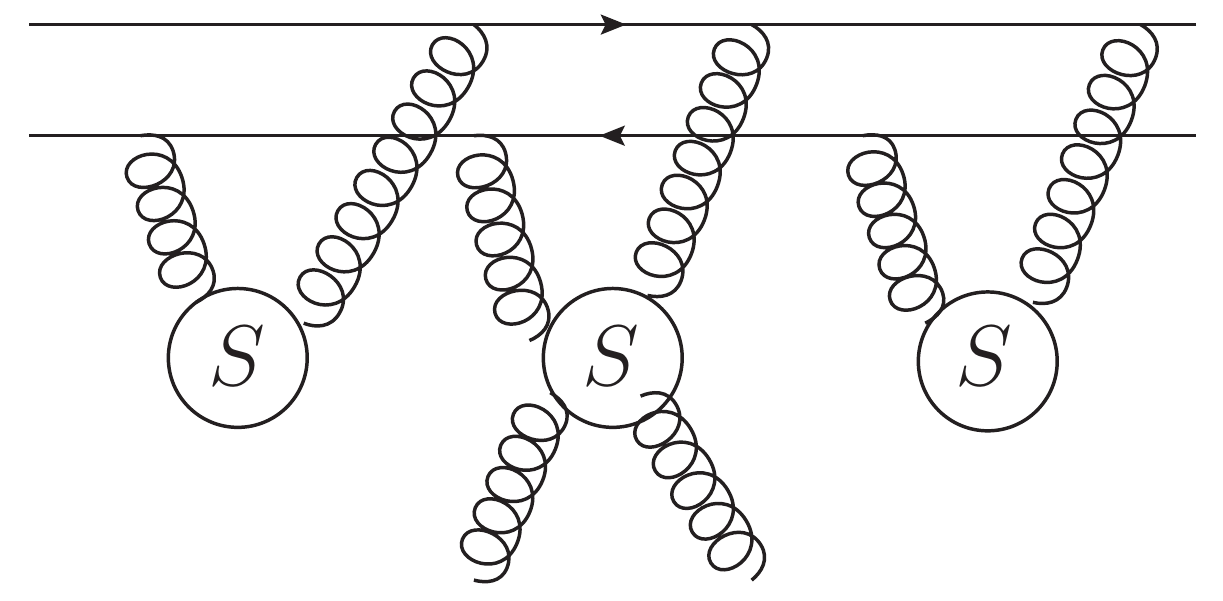}
\\
(a)&(b)
\end{tabular}
\caption{(a)Example of two particle reducible graphs in which some infrared gluons in the out most irreducible subgraph are exchanged between heavy quarks and other undetected energetic particles;(b)Example of two particle reducible graphs in which all infrared gluons in the out most irreducible subgraph are exchanged between the heavy quark pair. }
\label{fig:2Pr}
\end{figure}

We then consider two particle reducible graphs. Examples of these diagrams are shown in Fig.\ref{fig:2Pr}.
For diagrams in which all infrared gluons in the out most irreducible subgraph are exchanged between the heavy quark pair, like the second diagram in Fig.\ref{fig:2Pr}, contributions of the out most irreducible subgraph can be absorbed into wave functions of bound states of the heavy quark pair. After this absorption, contributions of these diagrams can be factorized into two parts. The first part corresponding to diagrams in which some infrared gluons in the out most irreducible subgraph are exchanged between heavy quarks and other undetected energetic particles. The second part corresponds to the evolution of the heavy quark pair into bound states.

For diagrams in which some infrared gluons in the out most irreducible subgraph are exchanged between heavy quarks and other undetected energetic particles, like the first diagram in Fig.\ref{fig:2Pr}, we can deform the integral path so that momenta of gluons exchanged between heavy quarks and other undetected energetic particles are far from Coulomb region. We denote the momenta scale of gluons in the  in the outmost subgraph as $|\vec{q}_{1}|$. After the deformation, momenta scale of gluons in other irreducible subgraphs is no less than $|\vec{q}_{1}|$ on leading pinch singular surfaces. We can first absorb contributions of these irreducible subgraphs into the hard subgraph. Infrared divergences caused by gluons in  the outmost subgraph cancel out as proved in last subsection.

After the cancelation, we consider contributions of the out most and the secondary outer irreducible subgraphs, with momenta scale of gluons in these subgraphs denoted as $|\vec{q}_{2}|$. We notice that contributions of the momenta region in which momenta scale of gluons in the outmost irreducible subgraph is much smaller than that in the secondary outer irreducible subgraph cancel out as analysed in last paragraph.
We thus require that momenta scale of gluons in these two subgraphs equal to each other. We then take the proofs in Sec.\ref{GY:2PI} and Sec.\ref{cancellation:2PI } and see the cancellation of topologically unfactorized infrared divergences in these two subgraphs.      We repeat the procedure and see that topologically unfactorized infrared divergences do cancel out in two particle reducible graphs.

According to above proofs, we conclude that  topologically unfactorized infrared divergences cancel out in the process considered here at leading order in $\Gamma_{H}$.

\section{Conclusion}
\label{conc}

We present the detailed proof of cancellation of topologically unfactorized infrared divergences in inclusive production of lepton pair close to the heavy quarkonium mass at leading order in the width of the heavy qaurkonium.   Such cancellation is crucial in proof of NRQCD factorization in these processes. Once NRQCD factorization holds in these processes, we get an equivalent  NRQCD factorization theorem for inclusive production of heavy quarkonia at leading order in the width of heavy qaurkonia.

We do not consider effects of the electro-weak interactions here. For QED interactions, infrared divergences caused by photons coupled to the heavy quark pair cancel out according to the similar proof presented here for those of gluons. Couplings between photons and the detected lepton pair do not change the electric charge of leptons. Infrared divergences caused by such couplings cancel out according to the KLN  cancellation. The weak interaction may cause the instability of heavy quarks and make the problem more complicated. We thus neglect these effects here.

Cancellation of  topologically unfactorized infrared divergences in the process considered here rely on the fact that parent states of the lepton pair can be any thing that contain a heavy quark pair. Thus effects of transition between different bound states of  the heavy quark pair are taken into account here. This is in accordance with practical experiments as parent states of the lepton pair are not detected direct.

\section*{Acknowledgements}

The author thanks  Y. Q. Chen for for helpful discussions and important suggestions about the work.

\bibliography{cancellation}

\end{document}